\documentclass[12pt]{article}
\pdfoutput=1
\usepackage[utf8x]{inputenc}
\usepackage{amssymb}
\usepackage{graphicx}
\usepackage{hyperref}
\usepackage{color}
\usepackage{bigints}

\newcommand{\eq}{\begin{equation}}
\newcommand{\eqx}{\end{equation}}
\newcommand{\eqs}{\begin{equation*}}
\newcommand{\eqsx}{\end{equation*}}
\newcommand{\eqn}{\begin{eqnarray}}
\newcommand{\eqnx}{\end{eqnarray}}
\newcommand{\eqns}{\begin{eqnarray*}}
\newcommand{\eqnsx}{\end{eqnarray*}}

\newcommand{\f}[2]{\frac{#1}{#2}}

\newcommand{\cor}[1]{\left\langle{#1}\right\rangle}
\newcommand{\ket}[1]{\left|{#1}\right\rangle}

\newcommand{\HH}{{\mathcal H}}

\newcommand{\lm}{\lambda}

\newcommand{\Sg}{\Sigma}
\newcommand{\dl}{\delta}
\newcommand{\Dl}{\Delta}

\newcommand{\al}{\alpha}

\newcommand{\om}{\omega}
\newcommand{\Om}{\Omega}
\newcommand{\omz}{\omega_0}

\newcommand{\eps}{\varepsilon}
\newcommand{\qqqq}{\quad\quad\quad\quad}
\newcommand{\qq}{\quad\quad}

\newcommand{\nn}{{\cal N}}

\newcommand{\chib}{\bar{\chi}}
\newcommand{\Chi}{{\mathcal X}}
\newcommand{\FF}{{\mathcal F}}
\newcommand{\ff}{{\mathbf f}}
\newcommand{\SG}{\mathbf{\Sigma}}
\newcommand{\J}{\mathbf{J}}
\newcommand{\DD}{\mathcal{D}}
\newcommand{\PSI}{\mathbf{\Psi}}
\newcommand{\TF}{\mathbf{T_F}}

\newcommand{\dd}{\widetilde{D}}
\newcommand{\KK}{\mathcal{K}}
\newcommand{\jb}{\bar{j}}

\newcommand{\arr}[4]{\left(
\begin{matrix}
{#1} & {#2}\\
{#3} & {#4} \\
\end{matrix}
\right)}

\newcommand{\vech}[2]{\left(
\begin{matrix}
{#1} & {#2}\\
\end{matrix}
\right)}

\newcommand{\vecv}[2]{\left(
\begin{matrix}
{#1} \\
{#2} \\
\end{matrix}
\right)}

\newcommand{\oo}[1]{{\mathcal O}\left(#1\right)}

\usepackage{tikz}

\tikzstyle{tensor}=[rectangle,draw=blue!50,fill=blue!20,thick]
\tikzstyle{round}=[circle,draw=blue!50,fill=blue!20,thick]

\newcommand{\tikzname}[1]{#1}

\newcommand{\diagram}[1]{ \begin{array}{cc}\tikzname{\begin{tikzpicture}[scale=.5,every node/.style={sloped,allow upside down},baseline={([yshift=+0ex]current bounding box.center)},inner sep=1mm] #1 \end{tikzpicture}} \end{array} }

\newcommand{\drawTF}[1]{
\draw[tensor] #1 rectangle +(1.5,-5);
\path #1 +(0.75,-2.5) coordinate (A);
\draw (A) node {$T_F$};
}

\newcommand{\drawE}[1]{
\draw[tensor] #1 rectangle +(1.0,-5);
\path #1 +(0.5,-2.5) coordinate (A);
\draw (A) node {$E$};
}

\newcommand{\drawAOA}[2]{
\path #1 +(0,0) coordinate (A);
\path #1 +(0,-2) coordinate (B);
\path #1 +(0,-4) coordinate (C);
\node[tensor] (0) at (A) {$A^{\phantom{*}}$};
\node[tensor] (1) at (B) {#2};
\node[tensor] (2) at (C) {$A^*$};
\draw[-] (0) -- (1);
\draw[-] (1) -- (2);
}

\title{Exact bosonic Matrix Product States\\
(and holography)}

\author{Romuald A. Janik\thanks{e-mail: {\tt romuald.janik@gmail.com}} \\ \\ 
\small 
Institute of Physics\\\small
Jagiellonian University\\\small
ul. {\L}ojasiewicza 11\\\small 
30-348 Krak{\'o}w\\\small 
Poland}

\date{}

\begin{document}

\maketitle

\begin{abstract}
We derive an exact formula for a matrix product state (MPS) representation (or a PEPS in higher number of dimensions)
of the ground state of translationally invariant bosonic lattice systems
in terms of a single one-dimensional Euclidean quantum mechanical path integral with sources. We explicitly evaluate the general formula
in the special case of the one-dimensional Klein-Gordon harmonic chain, being a spatial discretization of 1+1 dimensional free boson QFT, 
obtaining an exact MPS with an infinite dimensional
bond space. We analytically diagonalize the transfer matrix obtaining two
Fock spaces with continuous modes and check that the exact MPS construction reproduces 
the correct correlation functions.
We also comment on possible holographic interpretations.
\end{abstract}

\vfill

\pagebreak

\section{Introduction}

Matrix Product States (MPS) and more general tensor network constructions arose from
the DMRG (Density Matrix Renormalization Group) program and are an effective way
of studying wavefunctions, spectra and correlation functions in various interacting
lattice systems \cite{DMRG}. 
Due to the large dimensionality of the Hilbert space of lattice systems,
generic wavefunctions would require an exponentially large number of parameters.
Tensor networks like MPS, their higher 
dimensional analogs -- Projected Entangled Pair States (PEPS) \cite{PEPS} and further refinements like 
Multiscale Entanglement Renormalization Ansatz (MERA) \cite{MERA} provide a very effective
variational ansatz with a much smaller number of parameters which captures
properties of low lying states of systems described by hamiltonians with local
interactions \cite{MPSGROUNDSTATE}. These ansatzae are build up from tensors with some legs in
the physical local Hilbert space at each site and some legs in some
auxiliary vector space of dimension $D$ -- the so-called bond space.
The coefficients of these tensors are found by numerically minimizing
the expectation value of the hamiltonian.
When one increases $D$ one gets better and better approximations, of course at
the cost of increasing computation time and storage requirements.

Apart from these practical applications, a new source of interest in 
tensor networks appeared coming from the AdS/CFT correspondence \cite{ADSCFT} or holography \cite{HOLOGRAPHY} -- the exact equivalence 
between nongravitational field theories\footnote{These field theories are often referred to in the context of holography as boundary theories.} and theories in a higher dimensional
spacetime with boundary which include gravity (predominantly superstring theories).
Despite immense progress and overwhelming quantitative evidence,
we do not have a microscopic understanding of the emergence of
the dual higher dimensional theories directly from the nongravitational
boundary field theory.

In \cite{SWINGLE}, Swingle first noticed a remarkable direct analogy between holographic computation
of entanglement entropy \cite{RYUTAKAYANAGI} and a corresponding computation within MERA.
Indeed there is an appealing similarity between the structure of MERA
and the geometry of Anti-de-Sitter spacetime relevant for holography
for conformal field theories.
In the case of continuous MERA (cMERA) \cite{cMERA}, a proposal was made how to
associate a concrete geometry (bulk metric) to the cMERA description
of the boundary field theory \cite{TAKAYANAGI}.
Since then, there has been an immense research effort in this direction (see e.g. \cite{HAPPY, OPTIM, RANDOMTENSOR}).
On a cautionary note, however, tensor network descriptions are also quite far from
what we would like to have in holography in certain important respects.
In particular, we do not have a natural bulk action nor any kind of 
field theoretical description (which could be potentially interpreted 
in terms of gravity coupled to a specific matter content).

In contrast to the practical applications of tensor networks as
a computational tool for finding wavefunctions and computing
correlation functions in concrete strongly interacting systems,
when we want to investigate their potential basis for holography,
we would like to have an \emph{exact} description which
generically necessitates an infinite dimensional auxiliary bond space.
Although variational MPS ansatzae with an infinite dimensional bond space have
already appeared in some contexts \cite{iMPS}, the present construction
has a different origin and aims at obtaining an exact description,
albeit for a relatively simple system.

Apart from the case of cMERA, there are very few cases of
exact tensor network wavefunctions for natural physical systems.
A notable example is the exact MPS for the XY spin chain \cite{exactXY},
which was the key source of inspiration for this work.
One should also mention a construction of successive approximate MPS
for representing CFT correlation functions \cite{MPScorrCFT}.

The main aim of the present paper is to construct an exact Matrix Product State
for a chain of bosons with harmonic interactions (\emph{aka} Klein-Gordon chain), which is
the standard discretization of a 1+1 free (massive or massless) boson.
This physically interesting (and trivially exactly solvable) system unfortunately does
not fall directly into the class of applicability of so-called Gaussian Matrix Product
States introduced in \cite{GAUSSIANMPS} (see comments in section VIIG of that paper), hence this investigation.
We hope that the exact Matrix Product State description will be useful for
testing hypothesis on the link of holography and tensor networks.

As a byproduct of the construction, we provide expressions 
(see formulas (\ref{e.exactgen})-(\ref{e.exactpeps}) below) for the exact MPS
(or higher dimensional PEPS) of bosonic lattice systems with a local 
self-interaction potential
in terms of a 1-dimensional Euclidean path integral with sources.
These formulas may hopefully be used as an alternative to numerical
variational approach for such systems. As the focus of this paper is different,
we do not perform any concrete experiments in this direction.

Let us finally comment on the rationale for choosing Matrix Product States
and the specific physical system as ingredients for the exact tensor network
construction.

On the one hand, we did not want to restrict to a purely conformal system, as scale invariance
imposes various very powerful symmetry constraints and a lot of freedom in e.g.
modifying the underlying metric as exploited in \cite{OPTIM} (but see also \cite{OPTIM2} for 
going beyond conformality). Nonconformal systems are, in a sense, much more rigid.
On the other hand, the Klein-Gordon chain has
of course a gapless limit (and one can get a conformal massless boson in the continuum
limit for an appropriate choice of parameters), but we would like to have a common description
with gapped systems.

The motivation for treating a discrete system instead of working directly in the continuum
despite computational complexities due to discreteness,
was that in this context the standard tensor networks would be directly applicable.
In addition, one would not have to deal with various infinities and their renormalization.
Moreover the exact formulas
may be potentially of interest for some condensed matter applications.

In the present paper we choose to work with Matrix Product States 
instead of MERA, which is typically the tensor network of choice in the context of potential 
holographic applications.
Matrix Product States are conceptually simpler, and in the case of large $D$ (bond dimension)
describe well both gapped and gapless systems. For MPS it is easy to implement exact
translation invariance even for a discrete system (and finite $D$). MERA, in contrast, 
introduces coarse-graining and the very structure of the tensor network does not respect
translation invariance by one lattice site.

Last but not least, the structure of a MPS is quite intriguing from the point of view
of holography. The boundary points (physical lattice sites) are
directly disconnected and interact only through the intermediate auxiliary bond space.
As for an exact MPS we expect the bond space to be infinite dimensional, it is 
tempting to speculate that the infinite dimensional auxiliary space could have
a natural interpretation in terms of space of functions in some emergent
dimension. The exploration of such a scenario was a direct motivation
for this work.

The plan of this paper is as follows. In section~2 we give some further holographic 
motivations for the computations carried out later in the paper. In section~3 we review
the formalism of Matrix Product States, the transfer matrix and the computation
of correlation functions. Then in section~4 we give the general construction of exact
MPS and PEPS for a wide class of bosonic systems. The following 3 sections
are devoted to implementing this construction for the Klein-Gordon chain,
diagonalizing the transfer matrix and computing correlation functions.
In section~8 we return to the holographic motivations and discuss the obtained results
in view of the points raised in section~2. We close the paper with a summary and outlook.

The reader interested just in the exact MPS constructions can safely omit 
sections 2 and 8. The reader interested in potential links with holography
can safely omit sections 6 and 7 and use section 5 just for setting the notation.

\section{Holographic motivation}
\label{s.motivholo}

As described in the introduction, one of the main motivations of the current work
is to try to identify some features characteristic of a holographic description
in a completely solvable example system for which one can find an \emph{exact} tensor
network description. 

A holographic description of a given $d$-dimensional physical system means that there exists
a higher dimensional theory (in at least $d+1$ dimensions with a boundary) which is equivalent to the
original one and thus allows us to compute observables like correlation functions,
entanglement entropy etc. using the higher dimensional theory. Moreover,
the computational holographic prescriptions typically have an interpretation of putting some objects/operator
insertions/boundary conditions at the boundary which defines the particular quantity
to compute in the higher dimensional theory.

A quintessential example is the original AdS/CFT correspondence linking
$\nn=4$ SYM and superstring theory in $AdS_5 \times S^5$ spacetime. The $AdS_5$ geometry
can be written in the following form
\eq
ds^2 = \f{\eta_{\mu\nu} dx^\mu dx^\nu +dz^2}{z^2}
\eqx
where $z \geq 0$ is the `holographic' coordinate, and $z=0$ is the boundary. 
The boundary is intuitively identified with the space where the original field theory 
lives. Note that the distance between any pair of points at the boundary is infinite,
so that all observables are computed through passage into the bulk. 
Intuitively this is quite similar to the description in terms of Matrix Product States (see next section for an introduction), where there is no direct connection between local Hilbert spaces at different physical sites ($i_k$ below), and all interactions occur through the auxiliary bond space (the space of $\chi_k$'s)
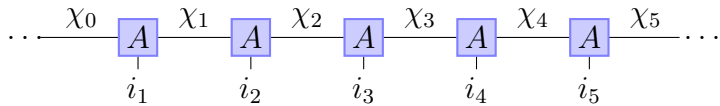
\[
\begin{tikzpicture}[inner sep=1mm,baseline={([yshift=0ex]current bounding box.center)}]

    \draw (0.2,0) -- (1.5,0);
    \draw (0,0) node {$\ldots$};
    \draw (0.75, 0.25) node {$\chi_0$};

    \draw (7.5,0) -- (8.7,0);
    \draw (9,0) node {$\ldots$};
    \draw (8.25, 0.25) node {$\chi_5$};

    \foreach \i in {1,...,5} {
    \pgfmathsetmacro{\ii}{1.5*\i}
        \node[tensor] (\i) at (\ii, 0) {$A$};
        \node (\i spin) at (\ii, -0.7) {$i_\i$};
        \draw[-] (\i) -- (\i spin);
    };
    \foreach \i in {1,...,4} {
        \pgfmathtruncatemacro{\iplusone}{\i + 1};
        \pgfmathsetmacro{\iiplusone}{1.5*\i + 0.75};
        \draw[-] (\i) -- (\iplusone);
        \draw (\iiplusone, 0.25) node {$\chi_\i$};
    };
\end{tikzpicture}
\]
A natural question which arises is whether the auxiliary bond space could have an
interpretation of an emergent dimension. The aim of constructing an \emph{exact}
Matrix Product State description in this paper was directly motivated by the above question.

\begin{figure}[t]
\[
\begin{tikzpicture}
\draw[very thick] (0,0) -- (10,0);
\draw (2,0) arc [start angle=180, end angle=0, radius=3];
\draw[fill=black] (2,0) circle [radius=0.1];
\draw[fill=black] (8,0) circle [radius=0.1];
\draw (2,-0.5) node {$\oo{0,t}$};
\draw (8,-0.5) node {$\oo{x,t}$};
\draw (10.75,0.1) node {$z=0$};
\draw [->] (9,1) -- (9,2);
\draw (9.25,1.5) node {$z$};
\end{tikzpicture}
\]
    \caption{The correlation function of an operator with large dimension can be calculated from the geodesic in the bulk between the two boundary points.}
    \label{f.adscorrfunc}
\end{figure}
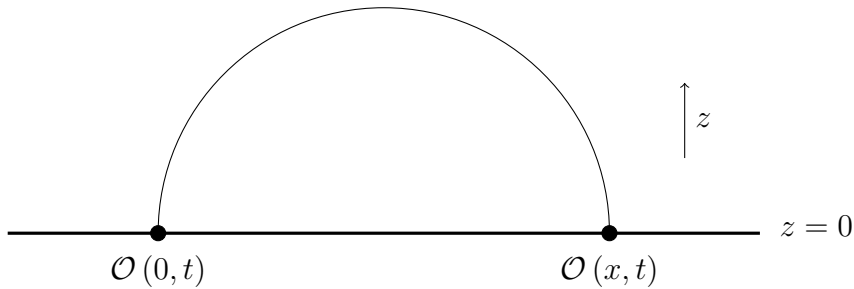

A simple example of a holographic prescription is the computation of
a correlation function of an operator of large dimension shown in Fig.~\ref{f.adscorrfunc}.
According to the AdS/CFT correspondence, we associate to each local operator 
in the field theory a corresponding bulk field. The two-point correlation function of the field theory operator
is roughly given\footnote{The exact prescription is slightly more involved, see \cite{GKP, WITTEN, FREEDMAN}.} by the boundary limit of the bulk Green's function of the corresponding bulk field \cite{ROUGH}
\eq
\cor{\oo{x,t}\oo{0,t}} = ``  \lim_{z \to 0 } z^\# \cor{\Phi(x,t,z) \Phi(0,t,z)} "
\eqx
When the bulk field is heavy, one can use the geodesic approximation for the Green's function, and one obtains the picture in Fig.~\ref{f.adscorrfunc}.

We will now specialize to the case of a 1+1 dimensional boundary and assume
that the bulk field has a free field mode expansion
\eq
\label{e.bulkmodes}
\Phi(x,t,z) = \int dk dp \,\mu(k,p) \left[ a^\dagger_{k,p} e^{i \Om(k,p)t-ipx} f_k(z) +
a_{k,p} e^{-i \Om(k,p)t+ipx} f_k^*(z) \right]
\eqx
The equal time correlation function would then have the following form
\eq
\label{e.bulkcorr}
\int dk dp  \,\mu(k,p) \left[\lim_{z \to 0 } z^\# |f_k(z)|^2\right]\, e^{i px} \equiv \int dk dp  \, F^2(k,p)\, e^{i px}
\eqx
with an unconstrained double integral reflecting 3-dimensional bulk modes.

Let us now contrast this with a direct evaluation of a 2-point correlation of a free
boundary theory\footnote{In general this is very far from generic examples of holography where the boundary theory is strongly interacting, but this will be relevant for
the system considered later in this paper.}, where the local operator $\oo{x,t}$ is just 
the local field $\phi(x,t)$ with the mode expansion
\eq
\phi(x,t)=\int dp\, \mu(p) \left[ a^\dagger_p e^{i \om(p)t-ipx} +
a_p e^{-i \om(p)t+ipx} \right]
\eqx
with $\mu(p) \propto 1/\sqrt{\om(p)}$. The resulting 2-point correlation function
is then given by a single integral
\eq
\int dp \, \mu^2(p) \, e^{ipx}
\eqx
Of course, for a theory on an integer lattice, the range of $p$ integration would be restricted to
the interval $p\in[-\pi,\pi]$.

The aim of this paper is to construct \emph{exact} Matrix Product States 
with an infinite dimensional auxiliary bond space 
for a bosonic lattice system and check whether
some auxiliary modes depending on $k$ appear in a natural way in the computation
of correlation functions using the Matrix Product State formalism.
In particular we would like to investigate whether the
result can be naturally recast into a formula of the form~(\ref{e.bulkcorr}).

\section{Matrix Product States}

Matrix Product States are a convenient ansatz for representing wavefunctions of 1D quantum mechanical systems which enables efficient computations using the variational method.
Their importance stems from the fact that
a generic wavefunction of a system with $L$ sites has an exponentially large number
of components $(dim\, \HH)^L$, where $\HH$ is the Hilbert space at each site. 
Even storing this data for large systems is problematic, let alone performing
a numerical optimization. Matrix Product States assume instead a factorized structure of the wavefunction
in terms of three-legged tensors, given below for a translationally invariant infinite system (see Fig.~\ref{f.mps}):
\eq
\Psi_{\ldots i_1i_2\ldots i_5\ldots} = \ldots A^{i_1}_{\chi_0\chi_1} A^{i_2}_{\chi_1\chi_2} A^{i_3}_{\chi_2\chi_3} \ldots
\eqx
The $i_k=1 \ldots dim\, \HH$ are physical indices, while $\chi_k$ and $\chi_{k+1}$  run from 1 to $D$, the dimension of an auxillary ``bond space''. The key ingredient is the 3 legged tensor
\eq
\label{e.Aiss}
\raisebox{-0.35cm}{$
\diagram{
\node[tensor] (0) at (2, 0) {$A$};
\node (1) at (0, 0) {$\chi$};
\node (2) at (4, 0) {$\chi'$};
\node (3) at (2, -1.5) {$i$};
\draw[-] (0) -- (1);
\draw[-] (0) -- (2);
\draw[-] (0) -- (3);
}$} = A^i_{\chi\chi'}
\eqx
and it is clear that the number of parameters does not grow exponentially with the system size.
As one increases $D$, one gets better and better approximation of the wavefunction. The approximation
is indeed very good for local hamiltonians with a gapped spectrum. However, by increasing $D$ one can
also get relatively good descriptions of gapless systems for intermediate distances.

\begin{figure}[t]

\[
\Psi_{\ldots i_1i_2\ldots i_5\ldots} = 
	\diagram{
	    \draw (-2.5,0) -- (-2.5,-1);
	    \draw (-2.5,-1.5) node {$\ldots$};
		\draw (-1.5,0) -- (-1.5,-1);
		\draw (-1.5,-1.5) node {$i_1$};
		\draw (-0.5,0) -- (-0.5,-1);
		\draw (-0.5,-1.5) node {$i_2$};
		\draw (0.5,0) -- (0.5,-1);
		\draw (0.5,-1.5) node {$i_3$};
		\draw (1.5,0) -- (1.5,-1);
		\draw (1.5,-1.5) node {$i_4$};
		\draw (2.5,0) -- (2.5,-1);
		\draw (2.5,-1.5) node {$i_5$};
	    \draw (3.5,0) -- (3.5,-1);
	    \draw (3.5,-1.5) node {$\ldots$};
		\draw[tensor] (-3,-1/2) rectangle (4,1/2);
		\draw (0.5,0) node {$\Psi$};
		
	}
\]

\[
\Psi_{\ldots i_1i_2\ldots i_5\ldots} = 
\begin{tikzpicture}[inner sep=1mm,baseline={([yshift=0ex]current bounding box.center)}]

    \draw (0.2,0) -- (1.5,0);
    \draw (0,0) node {$\ldots$};
    \draw (0.75, 0.25) node {$\chi_0$};

    \draw (7.5,0) -- (8.7,0);
    \draw (9,0) node {$\ldots$};
    \draw (8.25, 0.25) node {$\chi_5$};

    \foreach \i in {1,...,5} {
    \pgfmathsetmacro{\ii}{1.5*\i}
        \node[tensor] (\i) at (\ii, 0) {$A$};
        \node (\i spin) at (\ii, -0.7) {$i_\i$};
        \draw[-] (\i) -- (\i spin);
    };
    \foreach \i in {1,...,4} {
        \pgfmathtruncatemacro{\iplusone}{\i + 1};
        \pgfmathsetmacro{\iiplusone}{1.5*\i + 0.75};
        \draw[-] (\i) -- (\iplusone);
        \draw (\iiplusone, 0.25) node {$\chi_\i$};
    };
\end{tikzpicture}
\]

    \caption{Top: representation of the exact wavefunction of a translationally
    invariant system as an infinite component tensor.
    Bottom: a Matrix Product State (MPS) representation of the wavefunction in terms of an infinite product of
    identical tensors with 3 indices. The two auxillary bond indices $\chi_n$ and $\chi_{n+1}$ run in an auxillary vector space distinct from the physical Hilbert space at each site.}
    \label{f.mps}
\end{figure}

In higher number of dimensions, there is a direct analog of the MPS ansatz, called Projected Entangled Pair States (PEPS) where the basic building block (e.g. in 2D) is a 5 legged tensor
\eq
\label{e.Aissss}
\hspace{-0.5cm}\diagram{
\node[tensor] (0) at (2, 0) {$A$};
\node (1) at (0, 0) {$\chi$};
\node (2) at (4, 0) {$\chi'$};
\node (3) at (2, -1.5) {$\tilde{\chi}$};
\node (4) at (2,  1.5) {$\tilde{\chi}'$};
\node (5) at (4, -1.5) {$i$};
\draw[-] (0) -- (1);
\draw[-] (0) -- (2);
\draw[-] (0) -- (3);
\draw[-] (0) -- (4);
\draw[-] (0) -- (5);
} = A^i_{\chi\chi'\tilde{\chi}\tilde{\chi}'}
\eqx

The components of $A^i_{\chi\chi'}$ (or $A^i_{\chi\chi'\tilde{\chi}\tilde{\chi}'}$) are typically determined through numerical minimization of the expectation
value of the hamiltonian for given values of the bond dimension $D$. In this work we will concentrate,
however, on obtaining \emph{exact} formulas for the MPS/PEPS representation of bosonic lattice systems.
The key point is obtaining a natural identification of the infinite dimensional auxillary bond space.
Before we proceed with the construction, let us briefly review the other components of the MPS framework:
the transfer matrix and the computation of correlation functions.

In order to normalize the wave function and e.g. to compute correlation functions of local operators,
we need to contract the physical index of $A^i_{\chi\chi'}$ with the corresponding index of its complex conjugate.
The resulting composite tensor with 4 legs in bond space plays the role of the transfer matrix:
\eq
\label{e.tfgen}
\diagram{
\node[tensor] (0) at (3, 1) {$A^{\phantom{*}}$};
\node (1) at (1, 1) {$\chi$};
\node (2) at (5, 1) {$\chi'$};
\node[tensor] (3) at (3, -1) {$A^*$};
\node (4) at (1, -1) {$\bar{\chi}$};
\node (5) at (5, -1) {$\bar{\chi}'$};
\draw[-] (0) -- (1);
\draw[-] (0) -- (2);
\draw[-] (0) -- (3);
\draw[-] (3) -- (4);
\draw[-] (3) -- (5);

\draw (0,0) node {$=$};

\node (6) at (-1, 1) {$\chi'$};
\node (7) at (-5, 1) {$\chi$};
\node (8) at (-1, -1) {$\bar{\chi}'$};
\node (9) at (-5, -1) {$\bar{\chi}$};
\draw[-] (6) -- (7);
\draw[-] (8) -- (9);
\draw[tensor] (-3.75,1.5) rectangle (-2.25,-1.5);
\draw (-3,0) node {$T_F$};
}
\eqx
For correct normalizability on an infinite line, we require the existence of an eigenstate $\ket{E}$
with eigenvalue equal to unity $T_F \ket{E} = \ket{E}$ (which is also the highest eigenvalue of $T_F$). 
This can be represented pictorially as
\eq
\diagram{
\node (6) at (-0.5, 1) {};
\node (7) at (-5, 1) {$\chi$};
\node (8) at (-0.5, -1) {};
\node (9) at (-5, -1) {$\bar{\chi}$};
\draw[-] (6) -- (7);
\draw[-] (8) -- (9);
\draw[tensor] (-3.75,1.5) rectangle (-2.25,-1.5);
\draw (-3,0) node {$T_F$};
\draw[tensor] (-1, 1.5) rectangle (0, -1.5);
\draw (-0.5,0) node {$E$};

\draw (1,0) node {$=$};

\node (0) at (2, 1) {$\chi$};
\node (1) at (2, -1) {$\bar{\chi}$};
\node (2) at (4, 1) {};
\node (3) at (4, -1) {};
\draw[-] (0) -- (2);
\draw[-] (1) -- (3);

\draw[tensor] (3.5, 1.5) rectangle (4.5, -1.5);
\draw (4,0) node {$E$};
}
\eqx
With the above ingredients it is very simple to give a prescription for computing the correlation function
of two local operators on a line. It has the following obvious pictorial representation:
\eq
\label{e.corrfuncgen}
\diagram{
\draw (-3.5,0) -- (5,0);
\draw[dotted] (5,0) -- (6.5,0);
\draw (6.5,0) -- (12,0);

\draw (-3.5,-4) -- (5,-4);
\draw[dotted] (5,-4) -- (6.5,-4);
\draw (6.5,-4) -- (12,-4);

\drawE{(-4,0.5)}
\drawAOA{(-1.5,0.0)}{$O_0$}
\drawTF{(0, 0.5)}
\drawTF{(3, 0.5)}
\drawTF{(7, 0.5)}
\drawAOA{(10,0.0)}{$O_n$}
\drawE{(11.5,0.5)}

\draw[thin, rounded corners=8pt, dashed] (-5,1) rectangle (-0.5,-5);
\draw[thin, rounded corners=8pt, dashed] (9,1) rectangle (13.5,-5);

}
\eqx
In order to evaluate this expression one has to diagonalize the transfer matrix $T_F$ and 
compute the form factors of the individual operators, marked in the figure by the dashed lines.

In the rest of the paper we will carry out the construction of all these ingredients.
In section~\ref{s.mpsgeneral} we will provide general exact path integral expressions
for the MPS and PEPS states (\ref{e.Aiss}) and (\ref{e.Aissss}). In section~\ref{s.kgmps}
we will specialize to the case of the Klein-Gordon harmonic chain and explicitly evaluate the MPS formula, 
then we will proceed in section~\ref{s.transfer} to evaluate and diagonalize the transfer matrix
(\ref{e.tfgen}). As a cross check, we use the above results to evaluate the correlation
function (\ref{e.corrfuncgen}) in section~\ref{s.corrfunc}. We then try to look at
the obtained formulas for the correlation functions from a higher dimensional perspective in section~\ref{s.holoint}.

\section{General construction of exact bosonic MPS and PEPS}
\label{s.mpsgeneral}

Let us consider a general bosonic lattice system whose Lagrangian can be written in the form
\eq
L= \sum_n \f{M}{2} \dot{\phi}_n^2 - \f{D}{2} (\phi_{n+1} - \phi_n)^2  - V(\phi_n)
\eqx
or its say 2D version
\eq
L= \sum_{n,k} \f{M}{2} \dot{\phi}_{n,k}^2 - \f{D}{2} (\phi_{n+1,k} - \phi_{n,k})^2  - \f{D}{2} (\phi_{n,k+1} - \phi_{n,k})^2- V(\phi_{n,k})
\eqx
In this section we will allow the self interaction term $V(\phi)$ to be arbitrary, but in concrete calculations in the rest of the paper we will
specialize to the quadratic case $V(\phi)= \f{K}{2} \phi^2$.

We would like to construct an exact MPS/PEPS representation of the ground state wavefunction.
This is far from trivial even in the quadratic case where we have at our disposal 
an explicit formula for the exact wavefunction in position representation:
\eq
\label{e.psidirect}
\Psi(\ldots, \phi_{n-1}, \phi_n, \phi_{n+1}, \ldots) =e^{-\f{1}{2} \sum_{k,l} \phi_k C_{kl} \phi_l}
\eqx
where the (doubly infinite) matrix $C_{kl}$ is explicitly known\footnote{This case does not unfortunately
allow for a solution in terms of so-called Gaussian Matrix Product States introduced in \cite{GAUSSIANMPS},
which motivated the investigations in the present paper.}. Note that this form is very far from
an MPS ansatz, as in the wavefunction (\ref{e.psidirect}) all physical sites are directly coupled to each other. The problem of finding
a MPS representation corresponds to decoupling the physical sites through an intermediate auxillary bond space i.e. rewriting (\ref{e.psidirect}) as
\eq
\Psi(\ldots, \phi_{n-1}, \phi_n, \phi_{n+1}, \ldots) = \prod  \ldots A^{\phi_{n-1}}_{\chi_{n-1} \chi_n} 
A^{\phi_{n}}_{\chi_{n} \chi_{n+1}}
A^{\phi_{n-1}}_{\chi_{n+1} \chi_{n+2}} \ldots
\eqx

Let us now proceed with the construction in the generic case.
The ground state wavefunction up to an overall normalization can be of course obtained in the standard way through an Euclidean path integral  
\eq
\label{e.pathintgen}
\Psi(\ldots, \phi_{n-1}, \phi_n, \phi_{n+1}, \ldots) = \lim_{T\to \infty} \int_{\substack{u_n(0)=0\\ u_n(T)=\phi_n}} \DD u_n(\tau)
e^{-S_E[\ldots, u_{n-1}, u_n, u_{n+1}, \ldots]}
\eqx
with 
\eq
S_E = \sum_n \int_0^T \left( \f{M}{2} \dot{u}_n^2 + \f{D}{2} (u_{n+1} - u_n)^2  + V(u_n) \right) d\tau
\eqx
We would now like to rewrite this path integral expression as a product of MPS's. 
Intuitively, we should partition the Euclidean path integral 
into strips (or more concretely in the present discrete case
into lines ending at the physical sites) and reinterpret
these ingredients as successive MPS's. However in order to
do that we have to explicitly identify the bond variables.
At first glance, it would seem that one could just use the
Euclidean $u_n(\tau)$ in this role. However it is difficult,
if not impossible, to go along with this choice. Firstly, the
$u_n(\tau)$ are directly linked to the \emph{physical variables}
$\phi_n$ through the boundary condition $u_n(T)=\phi_n$.
In contrast, for discrete MPS's $A^i_{\chi\chi'}$ the bond space 
indices $\chi$, $\chi'$ are not constrained in any way by
the physical index $i$. Secondly, due to the direct coupling of
$u_n(\tau)$ and $u_{n\pm 1}(\tau)$ in the path integral formula
it is difficult to make a consistent choice (valid at
each lattice site) which of the three above variables
would be identified with the bond variables $\chi$ and $\chi'$.

In order to overcome these difficulties, we use a less
direct construction of the bond variables.
As a first step, we will rewrite the spatial interaction term in the action by introducing a functional Lagrange multiplier
\eq
e^{-\int_0^T \f{D}{2} (u_{n+1} - u_n)^2 d\tau}  = \int \DD s_n(\tau) \DD \chi_n(\tau) 
e^{-\int_0^T \left[ i \chi_n (s_n-(u_{n+1}-u_n)) +\f{D}{2} s_n^2 \right] d\tau}
\eqx
Now we will interchange the order of integration and first integrate out $s_n(\tau)$. Thus the above terms in the action will become
\eq
\label{e.chiterms}
\f{1}{2D} \chi_n^2 +i \chi_n (u_n-u_{n+1})
\eqx
Plugging this back into the Euclidean path integral we see that we can now directly interpret the expression as a product of
MPS's with the bond space being the space of the Lagrange multiplier functions $\chi_n(\tau)$ defined
on a half line $\tau \in [0,+\infty)$. Thus the conventional finite dimensional approximate MPS
is substituted by
\eq
A^i_{\chi \chi'}    \longrightarrow
A^\phi [\chi(\tau), \chi'(\tau)]
\eqx
and in the product of MPS's, the summation over the intermediate bond space index becomes a functional integral
\eq
\ldots \sum_{\chi'=1}^{D} A^{i_n}_{\chi \chi'} A^{i_{n+1}}_{\chi' \chi''} \ldots   \longrightarrow
\ldots \int \DD \chi'(\tau)  A^{\phi_n} [\chi(\tau), \chi'(\tau)] A^{\phi_{n+1}} [\chi'(\tau), \chi''(\tau)] \ldots
\eqx
The final formula for the exact MPS $A^\phi [\chi(\tau), \chi'(\tau)]$ is thus
\eq
\label{e.exactgen}
\raisebox{-0.35cm}{$
\diagram{
\node[tensor] (0) at (2, 0) {$A$};
\node (1) at (0, 0) {$\scriptstyle \chi(\tau)$};
\node (2) at (4, 0) {$\scriptstyle \chi'(\tau)$};
\node (3) at (2, -1.5) {$\scriptstyle \phi$};
\draw[-] (0) -- (1);
\draw[-] (0) -- (2);
\draw[-] (0) -- (3);
}$} = \lim_{T\to \infty} \int\displaylimits_{\substack{u(0)=0\\ u(T)=\phi}} \!\!\!\!\!\! \DD u(\tau)
e^{-\int_0^T d\tau \f{M}{2} \dot{u}^2 +V(u) +i (\chi-\chi')u +\f{1}{4D} \chi^2  +\f{1}{4D} {\chi'}^2}
\eqx
where we distributed the quadratic terms $\f{1}{2D} \chi_n^2$ of (\ref{e.chiterms}) symmetrically
into the two neighbouring MPS's. Clearly the same construction goes through
in higher number of dimensions providing an exact representation for PEPS, the only difference being additional $\chi$'s
in the other spatial directions.
\eq
\label{e.exactpeps}
\hspace{-0.5cm}\diagram{
\node[tensor] (0) at (2, 0) {$A$};
\node (1) at (0, 0) {$\scriptstyle \chi(\tau)$};
\node (2) at (4, 0) {$\scriptstyle \chi'(\tau)$};
\node (3) at (2, -1.5) {$\scriptstyle \tilde{\chi}(\tau)$};
\node (4) at (2,  1.5) {$\scriptstyle \tilde{\chi}'(\tau)$};
\node (5) at (4, -1.5) {$\scriptstyle \phi$};
\draw[-] (0) -- (1);
\draw[-] (0) -- (2);
\draw[-] (0) -- (3);
\draw[-] (0) -- (4);
\draw[-] (0) -- (5);
} \!\!= 
\lim_{T\to \infty} \!\!\!\! \int\displaylimits_{\substack{u(0)=0\\ u(T)=\phi}} \!\!\!\!\!\! \DD u(\tau)
e^{-\int_0^T d\tau \f{M}{2} \dot{u}^2 +V(u) +i (\chi-\chi'+\tilde{\chi}- \tilde{\chi}')u +
\f{1}{4D} (\chi^2  + {\chi'}^2 + \tilde{\chi}^2  + \mbox{$\scriptstyle\tilde{\chi}'$}^2)}
\eqx
Of course even for lattices in higher number of dimensions, the above Euclidean path integral 
remains just one-dimensional. 
Note that the formulas (\ref{e.exactgen})-(\ref{e.exactpeps}) are valid for an
arbitrary interaction potential $V(u)$.
The above formulas  may thus be used as an alternative to
the conventional variational approach by picking some
finite dimensional basis in the functional space of $\chi(\tau)$'s and obtaining matrix elements in that basis
by numerically evaluating the 1D Euclidean path integral. Alternatively one could perhaps employ various quantum mechanical
approximation schemes for obtaining approximate analytical results.
In particular, it should be feasible to apply perturbation theory
in their evaluation.

In the present paper we will from now on concentrate on the case of quadratic potential $V(u)= \f{1}{2} K u^2$ which corresponds
to a spatial discretization of Klein-Gordon theory in 1+1 dimensions. In this case one can evaluate explicitly the MPS (\ref{e.exactgen}),
as well as diagonalize the resulting transfer matrix.

\section{Exact MPS for the Klein-Gordon harmonic chain}
\label{s.kgmps}

The Klein-Gordon harmonic chain is given by the Lagrangian
\eq
L= \sum_n \f{M}{2} \dot{\phi}_n^2 - \f{D}{2} (\phi_{n+1} - \phi_n)^2  -  \f{K}{2} \phi_n^2
\eqx
In the following we will denote $\omz=\sqrt{K/M}$.
The above system is a standard spatial discretization of the free boson when one takes the lattice spacing $\eps \to 0$ with the following identifications
\eq
\label{e.continuum}
M=\eps \qq D=\f{1}{\eps} \qq K=m_0^2 \eps \qq \omz=m_0
\eqx
and $m_0$ is the mass in the continuum.
However all our subsequent computations will be valid in the discrete case for any values of the parameters $M$, $D$ and $K$.
We will just comment about what happens in the continuum limit.
For completeness let us recall that, in the general discrete case, the excitations of the harmonic chain 
have energies
\eq
\label{e.physdispersion}
E(p) = \sqrt{\f{K}{M} +\f{4D}{M} \sin^2 \f{p}{2} }
\eqx
with $p \in (-\pi,\pi)$.

As a cross check of the MPS construction and the diagonalization of the transfer matrix,
in this paper we will compute the correlation functions $\cor{\phi_0 \phi_m}$ and $\cor{\pi_0 \pi_m}$,
where $\pi_m = M \dot{\phi}_m$. These are given in the discrete case by quite involved expressions\footnote{Of course these are just Fourier transforms of the physical dispersion relation (\ref{e.physdispersion})
or its inverse, see e.g. (\ref{e.corphiphifourier}) later.}
which we quote here from 
\cite{CORRFUNCEXACT}.
\eq
\label{e.corr1}
\cor{\phi_0 \phi_m} = \f{1}{\sqrt{MK+2DM}} \cdot g_m^{(\infty)}(z)  \quad
\cor{\pi_0 \pi_m} = \sqrt{MK+2DM}\cdot h_m^{(\infty)}(z) 
\eqx
with
\eqn
g_m^{(\infty)}(z) &=&  \f{z^m}{2\mu} \binom{m-\f{1}{2}}{m} 
{}_2 F_1 \left( \f{1}{2}, m+ \f{1}{2}, m+1; z^2 \right) \\
h_m^{(\infty)}(z)  &=&  \f{\mu z^m}{2} \binom{m-\f{3}{2}}{m} 
{}_2 F_1 \left( -\f{1}{2}, m- \f{1}{2}, m+1; z^2 \right)
\label{e.corr4}
\eqnx
where
\eq
\mu = \f{1}{\sqrt{1+z^2}} \qq z=\f{1-\sqrt{1-\tilde{\al}^2}}{\tilde{\al}} \qq \tilde{\al} = \f{2D}{K+2D}
\eqx

The formula for the MPS (\ref{e.exactgen}) can be evaluated using the well known driven harmonic oscillator path integral. 
In view of diagonalizing the transfer matrix in the following section, it is convenient to keep $T$ in (\ref{e.exactgen}) finite as a regularization parameter.
The result for $A^\phi[\chi(\tau),\chi'(\tau)] $ is
\eq
\label{e.mpsexact}
e^{-\f{M}{2} \phi^2 \omz \coth \omz T -i \phi \int_0^T \f{\sinh \omz \tau}{\sinh \omz T} j(\tau) d\tau 
-\f{1}{2M} \int_0^T \int_0^T j(\tau) \Dl(\tau,\tau') j(\tau') d\tau d\tau'
-\f{1}{4D} \int_0^T \chi(\tau)^2+\chi'(\tau)^2 d\tau}
\eqx
where
\eq
j(\tau) = \chi(\tau) - \chi'(\tau)
\eqx
and
\eq
\Dl(\tau,\tau') = \f{\sinh \omz \tau_< \, \sinh \omz(T-\tau_>)}{\omz \sinh \omz T}
\eqx
We will fix the overall normalization when we discuss the transfer matrix in the following section.
We see that the dependence on $\phi$ is Gaussian while the dependence on the arguments of the MPS,
i.e. on $\chi(\tau)$ and $\chi'(\tau)$, although also Gaussian is nonlocal. Despite that, one
can explicitly diagonalize the transfer matrix, which we will do in the next section.

\section{The transfer matrix}
\label{s.transfer}

\begin{figure}
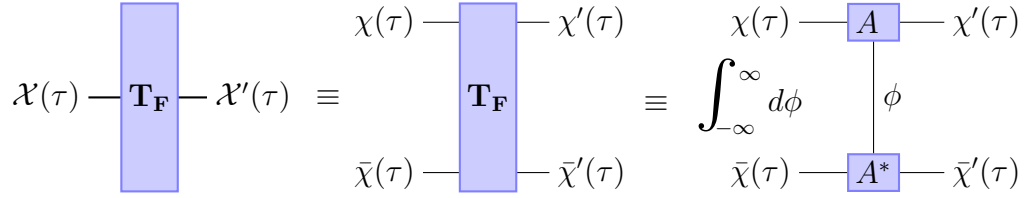

\[
\diagram{

\coordinate (B) at (-9,0.5);
\path (B) +(0.75,-2.5) coordinate (B0);
\path (B) +(-2,-2.5) coordinate (B1);
\path (B) +(3.5,-2.5) coordinate (B3);
\node (1) at (B1) {$\Chi(\tau)$};
\node (3) at (B3) {$\Chi'(\tau)$};
\draw[thick] (1) -- (3);
\draw[tensor] (B) rectangle +(1.5,-5);
\draw (B0) node {$\TF$};

\draw (-3.5,-2) node {$\equiv$};

\coordinate (A) at (0,0.5);
\path (A) +(0.75,-2.5) coordinate (A0);
\path (A) +(-2,-0.5) coordinate (A1);
\path (A) +(-2,-4.5) coordinate (A2);
\path (A) +(3.5,-0.5) coordinate (A3);
\path (A) +(3.5,-4.5) coordinate (A4);
\node (1) at (A1) {$\chi(\tau)$};
\node (2) at (A2) {$\chib(\tau)$};
\node (3) at (A3) {$\chi'(\tau)$};
\node (4) at (A4) {$\chib'(\tau)$};
\draw (1) -- (3);
\draw (2) -- (4);
\draw[tensor] (A) rectangle +(1.5,-5);
\draw (A0) node {$\TF$};

\draw (7,-2) node {$\equiv\;\; \bigint_{-\infty}^\infty d\phi $};

\coordinate (C) at (11,0);

\path (C) +(-3,0) coordinate (D1);
\path (C) +(-3,-4) coordinate (D2);
\path (C) +(3,0) coordinate (D3);
\path (C) +(3,-4) coordinate (D4);
\node (1) at (D1) {$\chi(\tau)$};
\node (2) at (D2) {$\chib(\tau)$};
\node (3) at (D3) {$\chi'(\tau)$};
\node (4) at (D4) {$\chib'(\tau)$};
\draw (1) -- (3);
\draw (2) -- (4);

\path (C) +(0,0) coordinate (C0);
\path (C) +(0.5,-2) coordinate (C1);
\path (C) +(0,-4) coordinate (C2);
\node[tensor] (5) at (C0) {$A^{\phantom{*}}$};
\node (6) at (C1) {$\phi$};
\node[tensor] (7) at (C2) {$A^*$};
\draw[-] (5) -- (7);

}
\]
\caption{Notations and definitions of the transfer matrix used in section~\ref{s.transfer}. \label{fig.tf}}
\end{figure}

Given the MPS state (\ref{e.mpsexact}) we construct the transfer matrix by performing the Gaussian integral over $\phi$ in
\eq
\TF[(\chi(\tau), \chib(\tau)), (\chi'(\tau), \chib'(\tau))] = \int_{-\infty}^{\infty} A^\phi [\chi(\tau), \chi'(\tau)] A^\phi [\chib(\tau), \chib'(\tau)]^* d\phi
\eqx
where $\chib(\tau)$ and $\chib'(\tau)$ denote the independent arguments of the lower MPS (with no relation to $\chi(\tau)$ and $\chi'(\tau)$ -- see
Fig.~\ref{fig.tf}). The ${}^*$
in the above denotes complex conjugation. This integral can be readily carried out giving
\eq
e^{-\f{1}{4M\omz \coth\omz T} \left( \int_0^T \f{\sinh \omz \tau}{\sinh \omz T} (j-\jb) d\tau \right)^2 - \f{1}{2M} \int_0^T\int_0^T (j \Dl j + \jb \Dl \jb)d\tau d\tau'
  -\f{1}{4D} \int_0^T \chi^2+\chi'^2+\chib^2+\chib'^2  d\tau }
\eqx

For ease of notation it is convenient to denote jointly the arguments of the upper and lower MPS entering the transfer matrix as
\eq
\Chi(\tau) = (\chi(\tau), \chib(\tau))  \qqqq 
\Chi'(\tau)= (\chi'(\tau), \chib'(\tau)) 
\eqx
We will also use the notation 
\eq
\al=\f{1}{2D}
\eqx
The transfer matrix can then be written as
\eq
\TF[\Chi(\tau), \Chi'(\tau)] = e^{-\f{1}{2}  \int \J(\tau) \SG(\tau,\tau') \J^T(\tau') d\tau d\tau'  -\f{\al}{2} \int \Chi(\tau)^2 + \Chi'(\tau)^2 d\tau }
\eqx
where
\eq
\J(\tau) = \Chi(\tau)-\Chi'(\tau)
\eqx
Diagonalization of the transfer matrix seems to be quite involved at first glance as we have to solve a \emph{functional} eigenvalue equation
involving a functional integral
\eq
\label{e.funceigen}
\int \DD \Chi'(\tau) \; \TF[\Chi(\tau), \Chi'(\tau)] \PSI[\Chi'(\tau)] = \Lambda\,  \PSI[\Chi(\tau)]
\eqx
This can be done, however, in two steps. Firstly, let us assume that the quadratic form $\SG(\tau,\tau')$ has been diagonalized
so that substituting
\eq
\label{e.diag1}
\Chi(\tau) = \sum_l x_l \ff_l(\tau) \qq \Chi'(\tau) = \sum_l x'_l \ff_l(\tau) \qq \text{with} \ \int_0^T \ff_i(\tau) \ff_j(\tau) = \dl_{ij}
\eqx
the transfer matrix becomes
\eq
\label{e.diag2}
\TF[\Chi(\tau), \Chi'(\tau)] = (normalization) \cdot  e^{-\f{1}{2} \sum_l \left( (x_l-x'_l) \Sg_l (x_l -x'_l) +\al x_l^2 +\al {x'_l}^2 \right)}
\eqx
where we introduced a normalization factor which should be chosen so that the MPS wavefunction is normalized to unity. This is translated
into the requirement that the highest eigenvalue of the transfer matrix should be equal to 1.

Solving the functional eigenvalue equation (\ref{e.funceigen}) now
factorizes into a product of 1D eigenvalue problems
\eq
\label{e.1dsubproblem}
(normalization) \cdot \int_{-\infty}^{\infty} dx' e^{-\f{1}{2}  (x-x')\Sg(x-x') -\f{\al}{2} x^2 -\f{\al}{2} {x'}^2 } \psi(x') = \lm \, \psi(x) 
\eqx

Each such 1D eigenvalue problem can be solved by harmonic oscillator wavefunctions
\eq
\label{e.dtildedef}
\psi_n(x) = \f{1}{\sqrt{2^n n!}} \left(\f{\dd}{\pi}\right)^{\f{1}{4}} H_n(\sqrt{\dd} x) e^{-\f{1}{2} \dd x^2}  \qq \text{where} \ \dd \equiv \sqrt{\al(\al+2\Sg)}
\eqx
with eigenvalue
\eq
\label{e.lambda}
\lm_n = \left( \f{\Sg}{\al+\Sg+\dd} \right)^n
\eqx
where we fixed the normalization factor in (\ref{e.1dsubproblem}) to
\eq
(normalization) = \sqrt{\f{\al+\Sg+\dd}{2\pi}}
\eqx
We thus see that we obtain in a natural way an underlying Fock space structure in the auxiliary bond space (which should not be confused with the physical excitations (\ref{e.physdispersion})).
To complete the picture, we need to determine the eigenvalues $\Sg_l$ and eigenfunctions $\ff_l(\tau)$ of the kernel $\SG(\tau,\tau')$ appearing in (\ref{e.diag1}) and (\ref{e.diag2}).

The $\SG(\tau,\tau')$ quadratic form takes the form
\eq
\J(\tau) \SG(\tau,\tau') \J^T(\tau') = \f{1}{M} \vech{j}{\jb}\arr{\f{\KK}{2} + \Dl}{-\f{\KK}{2}}{-\f{\KK}{2}}{\f{\KK}{2} + \Dl} \vecv{j}{\jb}
\eqx
where
\eqn
\KK &=& \f{1}{\omz \coth \omz T}\,  \f{\sinh \omz \tau}{\sinh \omz T}\, \f{\sinh \omz \tau'}{\sinh \omz T} \\
\Dl &=& \f{\sinh \omz \tau_< \, \sinh \omz(T-\tau_>)}{\omz \sinh \omz T}
\eqnx
The $2\times 2$ structure can be disentangled by passing to
\eq
j^\pm =j \pm \jb
\eqx
Then the quadratic form becomes
\eq
\f{1}{2M} j^+ \Dl j^+ + \f{1}{2M} j^- (\KK+\Dl) j^-
\eqx
We thus have two separate eigenvalue problems which we will consider in turn (we will now set $M=1$ and reinstate it at the end).

\subsubsection*{The $j^+$ modes}

Let us first consider
\eq
\label{e.eigeneq1}
\int_0^T \Dl(\tau,\tau') j^+(\tau') d\tau' = \Sg^+ j^+(\tau) 
\eqx
In order to solve it we use the fact that $\Dl(\tau,\tau')$ is a Green's function and satisfies
\eq
\left( \f{d^2}{d\tau^2} - \omz^2 \right) \Dl(\tau,\tau') = -\dl(\tau-\tau')
\eqx
Then acting with this operator on (\ref{e.eigeneq1}) leads to
\eq
\label{e.eqjplus}
\left( \f{d^2}{d\tau^2} - \omz^2 \right) j^+(\tau) = -\f{1}{\Sg^+} j^+(\tau)
\eqx
The fact that $\Dl(0,\tau')=\Dl(T,\tau')=0$ provides, again using (\ref{e.eigeneq1}),
the boundary conditions for equation (\ref{e.eqjplus})
\eq
j^+(0)=j^+(T)=0
\eqx
which leads to the following eigenfunctions and eigenvalues:
\eq
\ff^+_l(\tau) = \left( \f{1}{\sqrt{T}} \sin \f{l\pi\tau}{T}, \f{1}{\sqrt{T}} \sin \f{l\pi\tau}{T} \right)   \qq  
\Sg^+_l = \f{1}{M} \f{1}{\omz^2+l^2 \f{\pi^2}{T^2}}
\eqx

\subsubsection*{The $j^-$ modes}

We now turn to the second slightly more complicated problem. 
\eq
\label{e.eigeneq2}
\int_0^T \left[ \KK(\tau,\tau')+ \Dl(\tau,\tau')\right] j^-(\tau') d\tau' = \Sg^- j^-(\tau) 
\eqx
Since $\KK(\tau,\tau')$ is anihilated by $d/d\tau^2-\omz^2$, we will again get
\eq
\left( \f{d^2}{d\tau^2} - \omz^2 \right) j^-(\tau) = -\f{1}{\Sg^-} j^-(\tau)
\eqx
with $j^-(0)=0 $ but now supplemented with a more involved boundary condition at $\tau=T$:
\eq
\label{e.bcT}
\f{1}{\coth\omz T} \int_0^T  \f{\sinh \omz \tau'}{\sinh \omz T} \, j^-(\tau') d\tau' = \Sg^- j^-(T)
\eqx
The first two conditions determine $j^-(\tau) \propto \sin k\tau$ with $\Sg^- =1/(\omz^2+k^2)$.
Then the nontrivial condition (\ref{e.bcT}) simplifies to $\cos k T=0$. We thus get
\eq
\ff^-_l(\tau) = \left( \f{1}{\sqrt{T}} \sin  \f{\left(l+\f{1}{2}\right) \pi\tau}{T}, \f{-1}{\sqrt{T}} \sin  \f{\left(l+\f{1}{2}\right)\pi\tau}{T} \right)   \qq  
\Sg^-_l = \f{1}{M} \f{1}{\omz^2+ \f{\left(l+\f{1}{2}\right)^2 \pi^2}{T^2}}
\eqx

\subsubsection*{Transfer matrix eigenvalues}

Taking the $T\to \infty$ limit, the spectrum becomes continous
with $k=\f{(l+1/2)\pi}{T}$ or $k=\f{l\pi}{T}$, so in both cases we get
\eq
\label{e.sigmacont}
\Sg^\pm = \f{1}{M} \f{1}{\omz^2+k^2}
\eqx
We thus get two sets of modes parametrized by real positive $k$ which form a Fock space of particles\footnote{It is important to emphasize
that these particles live in the auxillary bond space of the transfer matrix and should not be confused
with the physical excitations of the harmonic chain.}.
The eigenvalue of the transfer matrix is given by
\eq
\prod_k  \left[ \lambda(k) \right]^{n(k)}
\eqx
where $n(k)$ is the integer particle number of mode $k$. $\lambda(k)$ is
the single particle eigenvalue (given by (\ref{e.lambda}) with $n=1$),
where we explicitly indicated its dependence on $k$. It is convenient
to write $\lambda(k)$ in exponential form
\eq
\lambda(k) \equiv e^{-E_{T_F}(k)}
\eqx
We will refer to the function $E_{T_F}(k)$ introduced here, as the transfer matrix dispersion relation.
Its explicit form is obtained by putting together (\ref{e.sigmacont}), the definition
of $\dd$ in (\ref{e.dtildedef}) and (\ref{e.lambda}).
It reads
\eq
\label{e.dispersion}
E_{T_F}(k) = \log \left(1+ x(k) + \sqrt{x(k) \left( 2+ x(k)\right)} \right)
\eqx
where
\eq
\label{e.alsg}
x(k) \equiv \f{\al}{\Sg} = \f{M}{2D} \left( \omz^2 +k^2 \right)
\eqx
To summarize, the eigenvalue of the transfer matrix is given by
\eq
\prod_k e^{-n(k) E_{T_F}(k)}
\eqx
where $n(k)$ is the integer particle number of mode $k$.
It is indeed quite surprising that the dispersion relation of the auxillary Fock space modes is so complicated for the simple bosonic chain.
In the continuum limit (\ref{e.continuum}) the dispersion relation (\ref{e.dispersion}) however significantly simplifies to
\eq
E_{T_F}(k) = \eps \sqrt{m_0^2 + k^2}
\eqx

\section{Correlation functions from the exact MPS construction}
\label{s.corrfunc}

We will now use the exact MPS and the results on the diagonalization
of the transfer matrix to compute the correlation functions
$\cor{\phi_0 \phi_m}$ and $\cor{\pi_0 \pi_m}$. Since their
exact expressions are quite complicated in the discrete case for
arbitrary separations (see (\ref{e.corr1})-(\ref{e.corr4})), 
this will be an important cross-check of our construction.

\subsubsection*{The $\cor{\phi_0 \phi_m}$ correlation function}

From the pictorial representation given in (\ref{e.corrfuncgen}),
we see that we have to evaluate the form factor of the operator $\phi$.
Consequently, at the site with operator insertion we have to compute
\eq
\label{e.AphiA}
\int_{-\infty}^{\infty}  A^\phi [\chib(\tau), \chib'(\tau)]^* 
\phi A^\phi [\chi(\tau), \chi'(\tau)] d\phi
\eqx
This is easily done with the result
\eq
\f{-i \J_-[\Chi(\tau), \Chi'(\tau)]}{2M \omz \coth \omz T} \cdot \TF[\Chi(\tau), \Chi'(\tau)] 
\eqx
where we use the notation
\eq
\label{e.Jpm}
\J_\pm[\Chi(\tau), \Chi'(\tau)] = \int_0^T \f{\sinh \omz \tau}{\sinh\omz T} j_\pm(\tau) d\tau
\eqx
and the product is just the ordinary multiplication.
Here for convenience we still keep the regulator $T$ finite. We will pass to the limit $T \to \infty$ at the end of the computation.
In the bond space we will use the basis which diagonalizes the transfer matrix
\eq
\Chi(\tau) = \sum_l x_l^+ \ff_l^+(\tau) + \sum_l x_l^- \ff_l^-(\tau) \qq \Chi'(\tau) = \sum_l {x'}_l^+ \ff_l^+(\tau) + \sum_l {x'}_l^- \ff_l^-(\tau)
\eqx
The matrix elements of $\TF$ then factorize into
\eq
\TF[\Chi(\tau), \Chi'(\tau)] = \prod_l K(x_l^+,{x'}_l^+) \prod_l K(x_l^-,{x'}_l^-)
\eqx 
with
\eq
K(x,x') = \sqrt{\f{\al+\Sg+\dd}{2\pi}} \cdot e^{-\f{1}{2}  (x-x')\Sg(x-x') -\f{\al}{2} x^2 -\f{\al}{2} {x'}^2 }
\eqx
The matrix elements of $\J_-$ determine the form factor of $\phi$
\eq
\J_-[\Chi(\tau), \Chi'(\tau)] = \f{2}{\sqrt{T}} \omz \coth \omz T \sum_{l=0}^\infty \f{(-1)^l}{\omz^2+ \left(l+\f{1}{2}\right)^2 \f{\pi^2}{T^2}} (x_l^- - {x'}_l^-)
\eqx
The wavefunction to the right of the rightmost operator $\phi$ is just the vacuum
\eq
\ket{E} = \prod_l \Psi_0(x_l^+) \prod_l \Psi_0(x_l^-)
\eqx
Since $\TF \ket{E} = \ket{E}$, we just have to evaluate
\eq
\int_{-\infty}^{\infty} (x-x') K(x,x') \Psi_0(x') dx'= \f{\al+\dd}{ \sqrt{2\dd} (\al+\Sg+\dd)} \Psi_1(x)
\eqx
Similarly we have
\eq
\int_{-\infty}^{\infty}  \Psi_0(x) (x-x') K(x,x') dx=  -\f{\al+\dd}{ \sqrt{2\dd} (\al+\Sg+\dd)} \Psi_1(x')
\eqx
Putting together the above equations as in (\ref{e.corrfuncgen}) leads to
\eq
\label{e.68}
\f{-1}{M^2 T} \sum_{l=0}^\infty \f{1}{\left( \omz^2+ \left(l+\f{1}{2}\right)^2 \f{\pi^2}{T^2}\right)^2}
\cdot
\left[ -\f{1}{2} \f{(\al+\dd)^2}{\dd (\al+\Sg+\dd)^2 } \right]
\cdot
\left( \f{\Sg}{\al+\Sg+\dd} \right)^{m-1}
\eqx
Removing the regulator by taking the limit $T \to \infty$ gives for the correlation function
\eq
\label{e.corphiphi}
\cor{\phi_0 \phi_m} = \int_0^\infty \f{dk}{2\pi} \underbrace{\f{(\al+\dd)^2}{\dd}}_{f_\phi(k)^2} 
\underbrace{\left( \f{\Sg}{\al+\Sg+\dd} \right)^{m+1}}_{e^{-(m+1) E_{T_F}(k)}}
\eqx
where $E_{T_F}(k)$ is given by (\ref{e.dispersion})-(\ref{e.alsg}), and the form factor of the $\phi$ field is
given by\footnote{Note that here the form factor is defined so that transfer matrix
eigenstates propagate for the full separation distance $m+1$ and not $m-1$ as in (\ref{e.corrfuncgen}).}
\eq
\label{e.fphik}
f_\phi(k) = \f{\al+\dd}{\sqrt{\dd}} =\sqrt{\al} \f{1+\sqrt{1+\f{2}{x(k)}}}{\left( 1+\f{2}{x(k)} \right)^{\f{1}{4}}}
\eqx
with $x(k)$ given by (\ref{e.alsg}). It is interesting to contrast the form of the correlation function obtained
here (\ref{e.corphiphi}) 
\eq
\cor{\phi_0 \phi_m} = \int_0^\infty \f{dk}{2\pi} f_\phi(k)^2 e^{-(m+1) E_{T_F}(k)}
\eqx
with the conventional expression in terms of the \emph{physical dispersion relation}
\eq
\label{e.corphiphifourier}
\cor{\phi_0 \phi_m} = \f{1}{2M} \int_{-\pi}^{\pi} \f{dp}{2\pi}  \f{e^{i p m}}{\sqrt{\f{K}{M} +\f{4D}{M} \sin^2 \f{p}{2} }}
\eqx
The modes appearing in the MPS expression should not be confused
with the physical modes in the first Brillouin zone $p\in (-\pi,\pi)$.
In addition they give an exponentially damped contribution to the
correlation function instead of an oscillatory one as in
the expression (\ref{e.corphiphifourier}).

The equivalence of the MPS expression (\ref{e.corphiphi}) with the
conventional one (\ref{e.corphiphifourier}) follows from the
considerations in section~\ref{s.holoint}.

\subsubsection*{The $\cor{\pi_0 \pi_m}$ correlation function}

The calculation of the correlation function $\cor{\pi_0 \pi_m}$
goes along the same lines with just some small modifications
which we underline here. 

Instead of (\ref{e.AphiA}) we now have to compute
\eqn
&& \int_{-\infty}^{\infty}  A^\phi [\chib(\tau), \chib'(\tau)]^* 
\pi A^\phi [\chi(\tau), \chi'(\tau)] d\phi = \nonumber\\
&&
\int_{-\infty}^{\infty}  A^\phi [\chib(\tau), \chib'(\tau)]^* 
\left( -i \f{\partial}{\partial \phi} \right) 
A^\phi [\chi(\tau), \chi'(\tau)] d\phi
\eqnx
The result is
\eq
-\f{1}{2} \J_+[\Chi(\tau), \Chi'(\tau)]  \TF[\Chi(\tau), \Chi'(\tau)] 
\eqx
with $\J_+$ defines as in (\ref{e.Jpm}). Its matrix elements are given by
\eq
\J_+[\Chi(\tau), \Chi'(\tau)] =  -2\pi T^{-\f{3}{2}} \sum_{l=1}^\infty 
\f{ l (-1)^l }{\omz^2+  \f{l^2\pi^2}{T^2}} (x_l^+ - {x'}_l^+)
\eqx
The difference w.r.t. the matrix elements of $\J_-$ comes from
the different boundary conditions of the $f^+_l$ and $f^-_l$
modes at $\tau=T$. Now we can immediately write the direct
analog of (\ref{e.68})
\eq
\f{1}{4} \sum_{l=1}^\infty \f{4l^2\pi^2 T^{-3}}{\left( \omz^2+ \f{l^2\pi^2}{T^2}\right)^2}
\cdot
\left[ -\f{1}{2} \f{(\al+\dd)^2}{\dd (\al+\Sg+\dd)^2 } \right]
\cdot
\left( \f{\Sg}{\al+\Sg+\dd} \right)^{m-1}
\eqx
and remove the regulator $T\to \infty$ to obtain the final result
\eq
\label{e.corpipi}
\cor{\pi_0 \pi_m} = \int_0^\infty \f{dk}{2\pi} \underbrace{\left[-M^2 k^2 \f{(\al+\dd)^2}{\dd}\right]}_{f_\pi(k)^2} 
\underbrace{\left( \f{\Sg}{\al+\Sg+\dd} \right)^{m+1}}_{e^{-(m+1) E_{T_F}(k)}}
\eqx
We checked numerically using Mathematica for a variety of parameters
that the obtained expressions (\ref{e.corphiphi}) and (\ref{e.corpipi}) 
agree with the exact answers given by (\ref{e.corr1})-(\ref{e.corr4}).

\section{Comments on holographic interpretation}
\label{s.holoint}

In the preceding sections we have constructed an exact MPS with an infinite dimensional bond
space for the bosonic Klein-Gordon harmonic chain. Moreover we have analytically diagonalized
the transfer matrix and uncovered a Fock space of two kinds of modes
each parametrized by a continuous parameter $k\in[0,+\infty)$. The eigenvalue of
the transfer matrix corresponding to both of these modes\footnote{More precisely for
single particle states in the corresponding Fock space.} is given by
\eq
\label{e.expminusetf}
e^{-E_{T_F}(k)} = \f{1}{1+ x(k) + \sqrt{x(k) \left( 2+ x(k)\right)}}
\eqx
with 
\eq
x(k) = \f{M}{2D} \left( \omz^2 +k^2 \right)
\eqx
In holography, the fields in the dual higher dimensional geometry
directly serve to compute correlation functions. This strongly suggests that the
potential tensor network counterpart should not really be the MPS but rather 
the MPS contracted with its complex conjugate. Hence at each lattice position on the boundary
we have effectively two Fock spaces with continuous modes parametrized by $k\geq 0$.
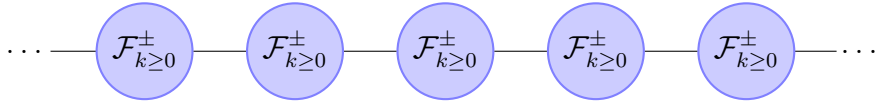
\[
\begin{tikzpicture}[inner sep=1mm,baseline={([yshift=0ex]current bounding box.center)}]

    \draw (0.75,0) -- (1.5,0);
    \draw (0.4,0) node {$\ldots$};

    \draw (10,0) -- (11.2,0);
    \draw (11.5,0) node {$\ldots$};

    \foreach \i in {1,...,5} {
    \pgfmathsetmacro{\ii}{2*\i}
        \node[round] (\i) at (\ii, 0) {$\FF_{k\geq 0}^\pm $};
    };
    \foreach \i in {1,...,4} {
        \pgfmathtruncatemacro{\iplusone}{\i + 1};
        \draw[-] (\i) -- (\iplusone);
    };
\end{tikzpicture}
\]
In the above, the Fock space modes are associated to each lattice site 
-- they are in position space w.r.t lattice.
In order to establish contact with the considerations in section~\ref{s.motivholo},
it is natural to pass to momentum space w.r.t lattice site positions.
Thus the new Fock space modes will be generated by
\eq
a^\dagger_{k,p},\, a_{k,p} \quad\text{and}\quad b^\dagger_{k,p},\, b_{k,p} \quad\quad 
\text{where} \quad k\geq 0, \, p\in[-\pi,\pi]
\eqx
From now on we just consider one set of these Fock spaces\footnote{More precisely the one with
nonvanishing form factor of the $\phi_n$ field.}.
Let us suppose, as in section~\ref{s.motivholo}, that a bulk field at $t=0$ and $z=0$
has the following mode expansion:
\eq
\Phi(n,t=0,z=0) = \int_0^\infty \f{dk}{2\pi} \int_{-\pi}^\pi \f{dp}{2\pi}  \, F(k,p)
\left[ a^\dagger_{k,p} e^{-ipn}  + a_{k,p} e^{ipn} \right]
\eqx
Then the 2-point equal time correlation function takes the form
\eq
\label{e.corrbulk}
\cor{\Phi(n,0,0)\Phi(0,0,0)} = \int_0^\infty \f{dk}{2\pi} \int_{-\pi}^\pi \f{dp}{2\pi}  \, F^2(k,p) e^{ipn}
\eqx
Let us compare this formula with the expression that we obtained from the MPS
computation
\eq
\label{e.corrmps}
\cor{\phi_n \phi_0} = \int_0^\infty \f{dk}{2\pi} f_\phi(k)^2 e^{-(|n|+1) E_{T_F}(k)}
\eqx
which we wrote now allowing for an arbitrary sign of $n$. Since the MPS computation was done
in position space w.r.t. the lattice sites, we should try to rewrite the result in momentum space.
Indeed we can use the formula
\eq
\int_{-\pi}^\pi \f{dp}{2\pi} \f{\sinh\eps}{\cosh\eps -\cos p} e^{ipn}  = e^{-|n|\eps}
\eqx
to find $F(k,p)$ so that (\ref{e.corrbulk}) and (\ref{e.corrmps}) exactly coincide.
This will occur when
\eq
F^2(k,p) = \f{\sinh E_{T_F}(k)}{\cosh E_{T_F}(k) -\cos p} f_\phi(k)^2 e^{- E_{T_F}(k)}
\eqx
Using the quite involved explicit formulas (\ref{e.fphik}) and (\ref{e.expminusetf}),
it is encouraging to find that after many cancellations the final result is
very simple and reads
\eq
\label{e.Fkp}
F^2(k,p) = \f{2\al}{1+x(k)-\cos p} = \f{1}{\f{M}{2} k^2+ 2D \sin^2 \f{p}{2} + \f{M}{2} \omz^2}
\eqx
Some comments are in order here. On the one hand, we see no trace of a 
nontrivial background geometry. What we recover is just the flat (discretized 
along the boundary direction) bulk corresponding the Euclidean path integral
(\ref{e.pathintgen}) as indicated by the form of (\ref{e.Fkp})
which looks exactly like the corresponding propagator.

On the other hand, we have a double unconstrained
integral which could be interpreted as coming from quantizing some
2+1 dimensional theory and reducing to a spatial timeslice
as in (\ref{e.bulkmodes})-(\ref{e.bulkcorr}) and not a single integral
characteristic of a quantized 1+1 dimensional theory.

Most probably these two interpretations are not in contradiction but
are rather two complementary points of view on the same expression.
In order to reach a definite conclusion, we need, however, to explicitly
study time evolution in the exact MPS framework as the picture must
be more involved than just (\ref{e.bulkmodes}) in order to account
for correlation functions of both $\phi$ and $\pi$ fields at different times
and correctly incorporate both sets of modes appearing in the diagonalization
of the transfer matrix.
We plan to address this issue in future work.

Finally, let us note that using the formulas in this section we can prove
analytically the equality of the MPS and standard formulas for the correlation
function $\cor{\phi_0\phi_n}$.
Indeed, performing first the integral over $p$ in (\ref{e.corrbulk})
with $F^2(k,p)$ given by the expression (\ref{e.Fkp}) reproduces, by construction, 
the MPS formula (\ref{e.corrmps}), while integrating first over $k$ yields the standard 
formula (\ref{e.corphiphifourier}) thus proving their equivalence.

\section{Summary and outlook}

The main results of this paper are the general formulas, valid for any
interaction potential, for exact Matrix Product States
or PEPS for a bosonic lattice system in terms of a 1-dimensional Euclidean path integral
with sources, and their evaluation for the 1-dimensional bosonic harmonic chain.
We hope that the general formulas may be of use as an alternative to the 
conventional variational approach for obtaining MPS and PEPS representations
by evaluating the exact formulas for some appropriately choosen subset of functions
from the bond space.

A very general feature of the above construction is that the auxiliary
bond space becomes the space of functions on a half line.
In the concrete case of the Klein-Gordon bosonic chain we can be much more explicit
and analytically diagonalize the transfer matrix.
Its eigenvalues and eigenfunctions form two Fock spaces
of modes parametrized by a continuous real parameter $k\in [0,+\infty)$.
The dispersion relation of these modes (understood here as minus the logarithm
of the transfer matrix for a single particle state) and given by (\ref{e.dispersion}) 
is surprisingly complicated for such an apparently simple physical system.
From the MPS computations of the 2-point equal time correlation functions
we obtained quite surprising novel integral representations of these
correlation functions which exactly reproduce the well known correlators
of this bosonic chain. It would be interesting to employ the MPS construction
to more nontrivial position space observables like entanglement entropy etc.

One of the main motivations for this work was that
an exact MPS representation of the ground state wavefunction of the Klein-Gordon
bosonic chain could be used as a testing ground for various hypothesis
on the relation of holography and tensor network constructions.
We hope that the result of this paper will be useful in this respect.

From the point of view of computing correlation functions,
the most natural counterpart of the holographic (spatial) geometry is not
a single MPS but rather the MPS contracted with its complex conjugate
keeping the intermediate bond space states arbitrary (i.e. not summed over).

Along these lines, we observed some encouraging and some discouraging results.
On the one hand, the infinite dimensional space of modes which diagonalize the transfer matrix
indeed can be identified with a Fock space of modes in some emergent dimension.
Also, the correlation functions can be written in a way reminiscent of a higher
dimensional quantized theory (although this is not the only possible interpretation).
On the other hand, we do not see a trace of some nontrivial background geometry.
So the relation of the obtained results to holographic expectations 
is not entirely clear-cut. 
However, we believe that it is necessary to analyze in detail temporal
evolution within this exact MPS framework to make any definite conclusions.
This remains as an important problem for future research.

Apart from that, there are numerous other interesting directions for further research,
like the development of approximate methods for treating the general formulas (\ref{e.exactgen})-(\ref{e.exactpeps})
in the interacting case,
possible applications of tensor network renormalization~\cite{TNRENORM} in the framework of the present construction, modifying
the boundary metric, moving directly to the continuum and merging the above approach
with some of the constructions in~\cite{QMHOLO}. We plan to investigate some of these issues
in future work.

\bigskip

\noindent{}{\bf Acknowledgments.} 
I would like to thank Marek Rams for lot of explanations on tensor networks, 
Michał Heller for very helpful remarks on the link to holography, Erik Tonni, 
Giuseppe Policastro and Leszek Hadasz for interesting discussions and comments.
I would like to thank Galileo Galilei Institute for Theoretical Physics for 
hospitality and the INFN for partial support as well as IPhT Saclay 
where part of this work was done.
This work was supported by NCN grant 2012/06/A/ST2/00396.

\end{document}